\documentclass[nohyper, a4paper,11pt,notoc]{article}

\usepackage{jheppub}
\usepackage{amsfonts,amsmath,amssymb}
\usepackage{bm,bbm,url}
\usepackage{graphicx,epsfig}
\usepackage{xspace}

\newcommand{\gsim}{\mathrel{\lower0.8ex\vbox{\lineskip=0.15ex\baselineskip=0ex
                   \hbox{$>$}\hbox{$\sim$}}}}
\newcommand{\lsim}{\mathrel{\lower0.8ex\vbox{\lineskip=0.15ex\baselineskip=0ex
                   \hbox{$<$}\hbox{$\sim$}}}}

\newcommand{\sla}[1]{{\raise.15ex\hbox{$/$}\kern-.57em #1}}
\newcommand{\Sla}[1]{\kern0.12em{\raise.15ex\hbox{$/$}\kern-.74em #1}}
\newcommand{\del}{\partial}
\newcommand{\tr}{{\rm Tr}}
\newcommand{\wbar}[1]{\overline{#1}}
\newcommand{\wtild}[1]{\widetilde{#1}}
\newcommand{\dubdel}{\!\!\stackrel{\leftrightarrow}{\raise.02ex\hbox{$\del$}}}
\newcommand{\dubD}{\!\!\stackrel{\leftrightarrow}{\raise.02ex\hbox{$D$}}}

\newcommand{\Gev}{\>{\rm GeV}}
\newcommand{\Tev}{\>{\rm TeV}}

\newcommand{\beq}{\begin{eqnarray}}
\newcommand{\eeq}{\end{eqnarray}}
\newcommand{\nn}{\nonumber}

\newcommand{\ep}{\epsilon}
\newcommand{\varep}{\varepsilon}

\newcommand{\SU}{{\rm SU}}

\newcommand{\col}{{\tilde{\rho}}}
\newcommand{\ourpi}{{\tilde{\pi}}}

\newcommand{\SUC}{\SU(3)_{\rm c}}

\newcommand{\Sherpa}{S\protect\scalebox{0.8}{HERPA}\xspace}
\newcommand{\Comix}{C\protect\scalebox{0.8}{OMIX}\xspace}
\newcommand{\Amegic}{A\protect\scalebox{0.8}{MEGIC++}\xspace}
\newcommand{\Delphes}{D\protect\scalebox{0.8}{ELPHES}\xspace}

\newcommand{\Fastjet}{F\protect\scalebox{0.8}{AST}J\protect\scalebox{0.8}{ET}\xspace}

\title{Hadronically decaying color-adjoint scalars at the LHC}

\author[a,b]{Steffen Schumann,}
\author[c]{Adrien Renaud,} 
\author[c]{Dirk Zerwas}
\affiliation[a]{Institut f\"ur Theoretische Physik, Universit\"at Heidelberg, 69120 Heidelberg, Germany}
\affiliation[b]{II. Physikalisches Institut, Universit\"at G\"ottingen, 37077 G\"ottingen, Germany}
\affiliation[c]{LAL, IN2P3/CNRS, Orsay, France}

\emailAdd{s.schumann@thphys.uni-heidelberg.de}
\emailAdd{renaud@lal.in2p3.fr}
\emailAdd{zerwas@lal.in2p3.fr}

\abstract{We study the phenomenology of the pair-production of scalar color-octet 
electroweak singlet states at the LHC. Such states appear in many extensions of the 
Standard Model. They can be pair-produced copiously at the LHC and will signal themselves 
as resonances in multijet final states. Beyond the QCD pair-production process we 
consider a vectorlike confinement scenario with an additional color-octet vector 
state. These vector particles can be produced in the s-channel and through their 
decay contribute to the scalar pair production. We point out the differences between the 
two hypotheses and device a strategy to distinguish them.}


\notoc
\begin{document}
\maketitle
\newpage

\section{Introduction}
\label{sec:intro}

The potential to produce new colored particles is one of the most prominent features 
of hadron colliders such as the Fermilab Tevatron or the CERN LHC. There, new-physics states that 
directly couple to the partonic initial state can be produced copiously for a wide range of masses. 
Most candidate theories for physics beyond the Standard Model (SM) predict the occurrence of such 
new particles and there is a continuous effort to search for hints for their production. Recent 
examples include the search for supersymmetric 
squarks and gluinos~\cite{Khachatryan:2011tk,daCosta:2011qk,Aad:2011hh,Chatrchyan:2011qs} or 
lepto-quarks~\cite{Khachatryan:2010mp,Aad:2011uv}. The various searches might be classified according 
to the considered final states. Besides a certain number of hadronic jets, remnants of the particle's 
color charge, this includes identified leptons from electroweak decays, associated photons and/or 
missing transverse energy from undetected decay products, e.g. a potential dark matter particle.

In this paper we analyse the pair-production process for new scalar color-octet states 
that are singlets under the electroweak gauge group. Such states appear in various extensions of the 
SM either as new fundamental particles \cite{Burdman:2006gy,Plehn:2008ae,Choi:2008ub} or composite 
objects \cite{Hill:2002ap,Kilic:2008ub}. As we consider states in the color-octet representation 
their QCD production cross section is fixed. The pair-production process, however, might have 
contributions depending on the theory that embeds the new scalars such as additional 
new particles that are produced resonantly in the $s$-channel and can decay into a pair of such octet 
scalars. The decays of the new scalar particles are of course highly model dependent and closely 
related to the underlying theory's spectrum. However, most likely, for a certain mass range two-body 
decays to SM partons will dominate and it is this decay mode we focus on here. As a consequence we 
have to search for the states under question in multijet final states, where they would signal 
themselves as dijet resonances. 

The precise signature for the hadronically decaying resonances certainly depends on the mass of the 
new states as well as the experimental trigger requirements. The continuum QCD production process will 
favour pair production at the kinematic threshold. In consequence we consider inclusive four-jet 
production as the search environment. In situations where the octets get produced with 
sufficiently large transverse momentum a dedicated subjet analysis in dijet final states might be 
the most powerful search strategy \cite{Bai:2011mr}. 

The central challenge of the analysis is to dig up the signal from the enormous QCD multijet 
background that exceeds the signal by orders of magnitude. The key point is to make optimal use 
of the kinematics of the signal events to represent the production of two {\em equal mass} narrow 
scalar resonances. In an earlier study the feasibility of this analysis was shown for the vectorlike 
confinement scenario \cite{Kilic:2008ub}. Here we further improve that analysis, extend it to 
discovering the pure QCD pair-production process, and point out how different models could be 
differentiated experimentally. We propose a data-driven approach to extract the relevant QCD 
multijet background, thereby minimizing the dependency on theoretical predictions.

In principle our candidate particles can also be produced singly at hadron colliders and 
there exist very stringent bounds from dijet resonance searches. For example 
ref.~\cite{ATLAS-CONF-2011-095} quotes a $95 \%$ CL exclusion limit of $1.91 \Tev$ for a 
color-octet scalar resonance with an order unity coupling to two gluons \cite{Han:2010rf}. 
However, as soon as this interaction is induced at the loop-level only, the single-production 
rate drops significantly thus evading the strong limit. Such scenarios we want to consider 
here, for which the pair-production process is the most promising discovery channel.

The paper is organized as follows. In Section~\ref{sec:models} we define our two benchmark 
scenarios, namely the most simple vectorlike confinement model and the SM extension by a 
complex scalar gluon that decays into two gluons with probability one. In 
Section~\ref{sec:sb_generation} we describe the tools used to simulate signal and background 
events. In Section~\ref{sec:evt_selection} we give details on our event-selection criteria 
and discuss a method that allows an extraction of the QCD background from data. In 
Section~\ref{sec:lhc_potential} we present results for our proposed search showing
how one could potentially distinguish between the two benchmark models. 
Section~\ref{sec:conclusions} contains our conclusions.

\section{Two benchmark models}
\label{sec:models}

Throughout this paper we want to consider two benchmark models for the production of electroweak 
singlet color-octet states. The first one is a vectorlike confinement scenario featuring besides 
our scalar candidate, the real octet hyper-pion, a further vector resonance in the color adjoint, 
called hyper-rho or coloron, that predominantly decays into a pair of hyper-pions. As the second 
template we consider the scalar partners of the gluino (called sgluons) as they appear in extended 
versions of the Minimal Supersymmetric Standard Model, such as the Minimal $R$-symmetric Supersymmetric 
Standard Model (MRSSM) \cite{Kribs:2007ac} or an $N=1/N=2$ hybrid model of supersymmetry with the $N=2$ 
supersymmetry implemented only in the gaugino sector \cite{Choi:2008ub}. In the following we will 
briefly define our two model hypotheses and discuss the different predictions for the pair-production 
rate of the two scalar octet candidates, the hyper-pion and the sgluon respectively.

\subsection{Hyper-pions in vectorlike confinement gauge theories}
\label{ssec:hyperpi}

The approach of vectorlike confining gauge theories at the TeV scale, first proposed in 
\cite{Kilic:2008pm} and later on generalized in \cite{Kilic:2009mi,Kilic:2010et}, provides 
an attractive ansatz for potential physics beyond the SM. Assuming a minimal extension of the SM
by new matter fields and making use of a fundamental mechanism already realized in nature, 
namely a confining strong force, a very rich phenomenology can be obtained from those theories 
described through rather few parameters only. The starting point is a quite minimal extension 
of the SM by new fermions in vectorlike representations of the SM gauge groups. These new fermions, 
charged under the strong and/or electroweak gauge group, feel a new strong gauge force called 
hypercolor (HC) that confines at the TeV scale, just as QCD confines at 
scales ${\cal{O}}(100$ MeV$)$. The emerging hyper-hadrons can be produced at the LHC signalling 
themselves through their decays mediated by hypercolor or SM interactions. The LHC phenomenology of 
vectorlike confinement will be dominated by the production of the lightest scalar and vector states, 
namely the pseudo-Nambu-Goldstone bosons called hyper-pions $\ourpi$ and the counterpart of the QCD 
$\rho$ meson, the hyper-rho $\col$.

We want to examine the most minimal vectorlike-confinement model that has been studied before 
in refs.~\cite{Kilic:2008pm,Kilic:2008ub,Dicus:2010bm,Sayre:2011ed}. We consider three new massless 
vectorlike fermions charged under QCD and hypercolor only. The hypercolor gauge group is fixed to
$\SU(3)_{\rm HC}$ and we assign the fermions as bi-fundamentals of the $\SUC \otimes \SU(3)_{\rm HC}$ 
product group and singlets under the electroweak interactions. The lowest lying states below the 
hypercolor-confinement scale $\Lambda_{HC} \sim$ TeV are the color-adjoint pseudo-Nambu--Goldstone 
hyper-pions $\ourpi$ and the octet hyper-rho or coloron $\col$. For this scaled-up version of QCD 
their mass ratio is approximated by 
\beq
\frac{m_{\ourpi}}{m_\col} \simeq 0.3\,,
\eeq
with $m_{\col}\sim \Lambda_{HC}$. The effective Lagrangian describing the dynamics and interactions of
the hyper-pion and hyper-rho is given by \cite{Kilic:2008ub}

\beq
  {\cal L}_{\rm eff}^{\rm HC}
  &=& -\frac{1}{4} G^{a}_{\mu\nu} G^{a\mu\nu} -\frac{1}{4}\tilde\rho^a_{\mu\nu}\tilde\rho^{a{\mu\nu}} +\frac{m_{\col}^{2}}{2}\col^{a}_{\mu}\col^{a\mu} +\wbar{q} i\gamma^{\mu}\left(\partial_{\mu}+ig_{3}\left(G_{\mu}+\varep\col_{\mu}\right)\right)q\nn\\
  && +\frac{1}{2} (D_{\mu}\ourpi)^{a}(D^{\mu}\ourpi)^{a}-\frac{m_{\ourpi}^{2}}{2}\ourpi^{a}\ourpi^{a}-g_{\col\ourpi\ourpi}f^{abc}\col^{a}_{\mu}\ourpi^{b}\del^{\mu}\ourpi^{c}-\frac{3g_{3}^{2}}{16\pi^{2}f_{\ourpi}}\tr\bigl[\ourpi G_{\mu\nu}\wtild{G}^{\mu\nu}\bigr]\nn\\
  && +\xi\frac{2i\alpha_{s}\sqrt{3}}{m^{2}_{\col}}\ \tr\left(\col^{\ \mu}_{\nu}\left[G^{\ \nu}_{\sigma},G^{\ \sigma}_{\mu}\right]\right)+i\chi g_{3}\tr\left(G_{\mu\nu}\left[\col^{\mu},\col^{\nu}\right]\right)\label{eq:model-Lag}  \>.
\eeq

\begin{figure}[!t]
  \centering
  \includegraphics[width=6cm]{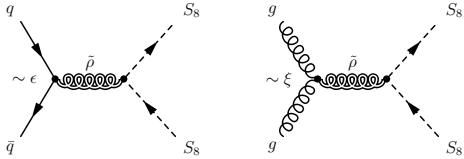}
  \caption{Feynman diagrams for resonant coloron production decaying into a 
    pair of color-adjoint scalars. Note that we consider $\ep=0.2$ 
    while we set $\xi=0$ such that the gluon-gluon fusion contribution vanishes.}
  \label{fig:Dia_Coloron_to_S8S8}
\end{figure}

The first line contains the SM and hyper-rho kinetic terms and the coupling of the hyper-rho to quarks
induced by kinetic mixing of the gluon and coloron fields. The second 
line summarises the kinetic and mass terms for the hyper-pion and the effective vertices that describe 
the decays $\col\to \ourpi\ourpi$ and $\ourpi\to gg$. The terms in the third line represent strong 
interaction matrix elements of the underlying theory that cannot be extracted from the QCD analog 
\cite{Kilic:2008ub}. The parameter $\xi$ is an undetermined number of order one. We use the most 
conservative assumption of $\xi=0$ as nonzero values enhance the pion pair-production cross section, 
cf. ref.~\cite{Kilic:2008ub}, through an additional resonant coloron contribution in the gluon-gluon 
fusion channel, cf. Figure~\ref{fig:Dia_Coloron_to_S8S8}. The term proportional to $\chi$ is of no 
relevance for our analysis as it contributes to the coloron pair-production process only \cite{Kilic:2008ub}. 

All other parameters of the model are fixed by just scaling up the QCD spectrum and the corresponding 
couplings of the low-energy QCD chiral Lagrangian \cite{Kilic:2008pm}. The mixing parameter $\ep$ is 
determined to $\ep\simeq 0.2$ and the coloron to hyper-pion coupling is given by $g_{\col\ourpi\ourpi}\simeq 6$.
The hyper-pion decay constant is determined through
\beq
\frac{f_\ourpi}{\Lambda_{\rm HC}} \simeq \frac{f_{\pi}}{\Lambda_{\rm QCD}}\,,
\eeq
accordingly the hyper-pion decays promptly into two gluons. The coloron is rather 
broad state, 
\beq
\frac{\Gamma_\col}{m_\col} \approx 0.19\,.
\eeq
Its dominant decay mode however is not into dijets but rather a pair of hyper-pions, determined by
\beq
\frac{\Gamma_{\col\to q\bar q}}{\Gamma_{\col\to\ourpi\ourpi}} \sim \frac{\ep}{g_{\col\ourpi\ourpi}}\,.
\eeq
\begin{figure}[!t]
  \centering
  \includegraphics[width=2.5cm]{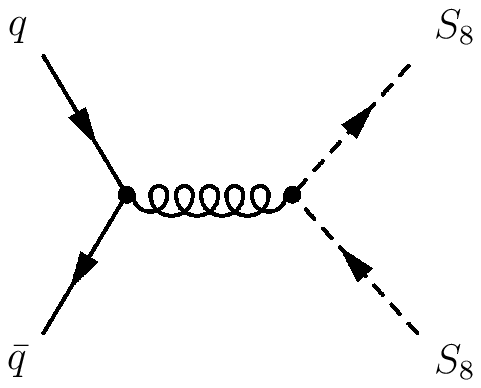}
  \hspace*{1cm}
  \includegraphics[width=6cm]{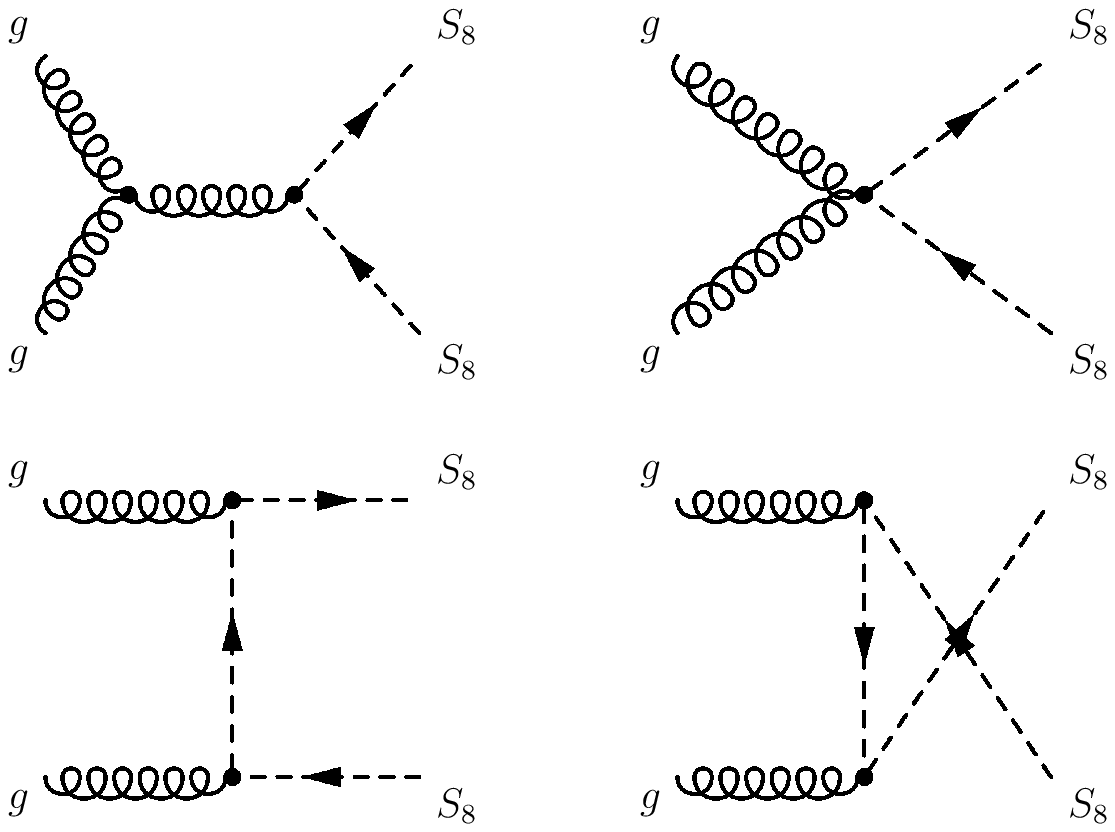}
  \caption{Feynman diagrams for the QCD pair production of color adjoints in the $q\bar q$ and $gg$ channel.}
  \label{fig:Diag_QCD_to_S8S8}
\end{figure}

\subsection{Sgluons in supersymmetric models with Dirac gauginos}
\label{ssec:sgluon}

Supersymmetric models that accomplish Dirac gauginos provide an attractive solution to the 
flavour problem of the MSSM \cite{Kribs:2007ac}. This is achieved through an additional 
suppression of dangerous flavour-violating processes. Considering a Dirac gluino, in order to 
account for its four degrees of freedom, besides the gluon two additional bosonic degrees of 
freedom must be added in a viable supersymmetric theory. This complex color-adjoint scalar, 
known as the sgluon, has explicit realisations in the context of the MRSSM \cite{Plehn:2008ae} 
or a phenomenological $N=1/N=2$ SUSY hybrid model as considered in \cite{Choi:2008ub}.

The sgluon has a gauge coupling to gluons and its pair-production rate is thus given by QCD. It 
interacts with quarks only at the one-loop level with the couplings being proportional to the 
quark masses. The sgluon-gluon-gluon interaction also occurs at the one-loop level only and is 
mediated by squarks. Accordingly resonant single-sgluon production is strongly suppressed. The 
sgluon coupling to two gluinos is the supersymmetric partner of the gluon coupling, while the 
coupling to squarks originates from $D$ terms and is proportional to the Dirac gluino 
mass \cite{Plehn:2008ae}.

The best laboratory to discover the sgluon is its pair-production process. In \cite{Plehn:2008ae} 
two split sgluon states have been considered. The work focused on the spectacular decay 
of the sgluon into top or anti-top accompanied by a light-flavour quark what leads to a 
jet-associated same-sign top signature. If heavy enough the sgluon can decay at the tree-level 
into pairs of gluinos or squarks. It is those decays that ref.~\cite{Choi:2008ub} 
focused on along with the decay into $t\bar t$, however, the authors considered a degenerate 
pair of sgluons.

In both scenarios the sgluon has a sizeable branching fraction into gluons that dominates 
certainly below the top threshold. It is this part of the sgluon phenomenology we want to study 
here. To be precise, we consider two degenerate sgluon states as in ref.~\cite{Choi:2008ub} and 
fix the sgluon branching ratio to gluons to unity independent of its actual mass. Such we 
simplify the underlying extended supersymmetric model but we still capture a large part of 
its phenomenology. In order to establish the supersymmetric theory that embeds the scalar 
adjoint the decay channels discussed in refs.~\cite{Plehn:2008ae,Choi:2008ub} need to be studied.

\begin{figure}[!t]
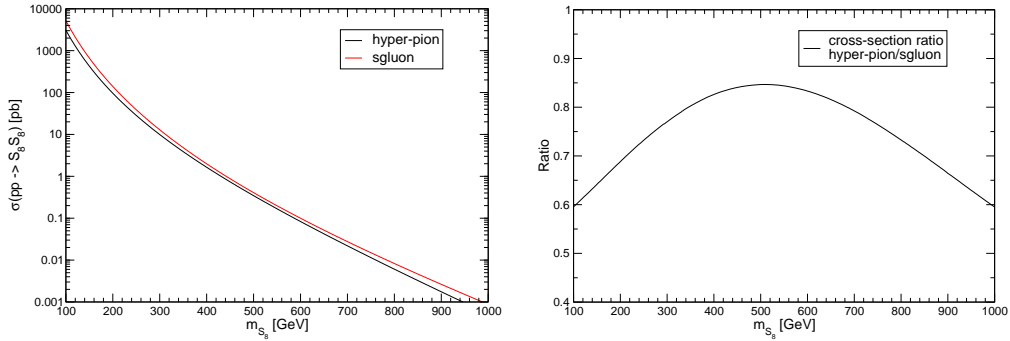

  \centering
  \includegraphics[width=6.5cm]{fig/sigma_SS.eps}
  \hspace*{2mm}
  \includegraphics[width=6.2cm]{fig/sigma_SS_ratio.eps}
  \caption{(Left) The tree-level total cross sections at the $7$ TeV LHC for the production of a 
    pair of octet scalars. Besides the hyper-pion signal we present the cross section for 
    producing a pair of scalar gluons \cite{Plehn:2008ae,Choi:2008ub}. (Right) The ratio of the 
    hyper-pion and sgluon cross section as function of the scalar mass.}
  \label{fig:Sigma_ScalarPairs}
\end{figure}

\subsection{Octet-scalar pair production}
\label{ssec:modelcomparison}

In Figure~\ref{fig:Diag_QCD_to_S8S8} we depict the Feynman diagrams for the QCD pair production
of hyper-pions and sgluons from $q\bar q$ and $gg$ initial states. The way we set up our sgluon 
benchmark model its pair-production rate exceeds the continuum QCD production rate for a pair of 
hyper-pions by a factor of two, consequence of the degeneracy of the two sgluon states. However, 
the actual hyper-pion pair-production rate is determined by the additional $s$-channel coloron 
contributions depicted in Figure~\ref{fig:Dia_Coloron_to_S8S8}. 
In Figure~\ref{fig:Sigma_ScalarPairs} (Left) we compare the $7 \Tev$ LHC cross sections for the 
hyper-pion and the sgluon pair-production signal as a function of the scalar mass. The ratio of 
the two cross sections is shown in Figure~\ref{fig:Sigma_ScalarPairs} (Right) as function of the 
mass. For low scalar masses the ratio is almost~$0.6$ as expected for a real scalar with respect 
to a complex scalar field. At higher masses the ratio approaches~$0.83$ as the contribution of 
the $s$-channel coloron exchange becomes more important before dropping back to $0.6$ at $1$~TeV.

The two models outlined above shall serve as concrete realisations of new color-adjoint 
scalars. As mentioned before such particles appear in many Standard Model extensions. 
The pair production of electroweak singlets decaying into heavy quarks has for example 
been considered in refs.~\cite{Chivukula:1991zk,Dobrescu:2007yp,Bai:2010dj}, rare decays 
have been studied in ref.~\cite{Zerwekh:2008mn}, phenomenological analyses of 
electroweak-charged adjoint scalar production can be found in 
refs.~\cite{Manohar:2006ga,Gerbush:2007fe,Bai:2010mn}.  

\section{Simulation of signal and background events}
\label{sec:sb_generation}

Before turning into the discussion of our search strategy and signal 
selection criteria we want to briefly describe the simulation tools employed 
to model the signal and background processes. 

We use the \Sherpa event generator \cite{Gleisberg:2008ta} to simulate
both, the hyper-pion and sgluon pair-production signal as well as the QCD 
multijet background. The signal models have been implemented in the 
matrix-element generator \Amegic \cite{Krauss:2001iv}. The Feynman rules
for the hyper-color model have been derived from Eq.~(\ref{eq:model-Lag}). 
The interactions relevant for the pair-production of sgluons are given by 
QCD and we assume the sgluon branching ratio to gluons equal to unity.

For the signal events we set the factorisation and renormalisation scales 
equal to the scalars mass, i.e. $\mu_F=\mu_R=m^2_{S_8}$. 
The decay of the scalar resonances into gluons we treat in the 
narrow-width-approximation, whereas we fully take into account off-shell 
effects for the broad coloron $s$-channel contribution in the hyper-color 
scenario. The parton-level signal events then get passed to \Sherpa's 
parton-shower and hadronisation routines 
\cite{Schumann:2007mg,Gleisberg:2008ta} in order to obtain realistic 
particle-level events.
 
The modelling of the QCD multijet backgrounds is a severe challenge 
\cite{Buckley:2011ms}. Up to now the next-to-leading order corrections to 
four-jet production are unknown so we have to rely on leading-order predictions. 
To accurately simulate multijet production we employ matrix-element parton-shower 
matching as implemented in \Sherpa \cite{Hoeche:2009rj,Hoeche:2009xc}. We consider 
the complete sets of tree-level matrix elements with up to six final-state partons 
from \Sherpa's matrix-element generator \Comix \cite{Gleisberg:2008fv}. The 
matrix elements of varying multiplicity get consistently combined to an inclusive 
sample of fully exclusive events through subsequent parton-shower evolution 
\cite{Schumann:2007mg} before subject to hadronisation. The renormalisation and 
factorisation scales are dynamically determined on an 
event-by-event basis according to the matching algorithm \cite{Hoeche:2009rj}.  
For the parton-separation parameter of the matching procedure we use 
$Q_{\rm cut} = 30 \Gev$. Note that \Sherpa's model for QCD multijet production has 
already successfully been validated against data from HERA \cite{Carli:2010cg}, 
Tevatron \cite{d0_r32,Abazov:2011rd} and the LHC \cite{daCosta:2011ni,Collaboration:2011tq}, 
what makes us confident to use a realistic background estimate throughout this study.
In Sec.~\ref{sec:evt_selection} we describe a strategy to extract the normalisation 
and shape of the QCD background in a data-driven approach.

To account for detector-resolution effects such as finite granularity and 
jet smearing we process the particle-level signal- and background events 
through the publically available detector-simulation package \Delphes 
\cite{Ovyn:2009tx}. \Delphes yields a reliable simulation for a prototypical 
LHC detector that has been used in several physics studies such 
as~\cite{deVisscher:2009zb,Ovyn:2008tm,Desch:2010gi}. For defining jets we make 
use of  the \Fastjet \cite{Cacciari:2005hq} package.

\section{Event selection and background estimation}
\label{sec:evt_selection}

Throughout the analysis we use anti-$k_T$ jets \cite{Cacciari:2008gp} with an $R$ 
parameter of $0.6$. Events are required to have at least four jets with a minimal $p_T$ 
of $50$\% of the mass of the resonance. Once this $p_T$ selection is applied the 
signal is composed of octet scalars produced with a non-negligible transverse boost. In 
the $\eta-\phi$ plane the separation of the gluon jets from the decay is typically 
$\Delta R\approx1$. We then consider all the possible pairings between the four 
highest-$p_T$ jets and retain the combination that minimizes 
$|\Delta R_{ij} -1|  + | \Delta R_{kl}  - 1|$ where $i,j,k,l$ denote the four leading jets. 
If one of the reconstructed resonances has a jet separation larger than $\Delta R_{ij} =1.6 $, 
the event is rejected. 

\begin{figure}[!t]
  \centering
  \includegraphics[width=7cm,height=6cm]{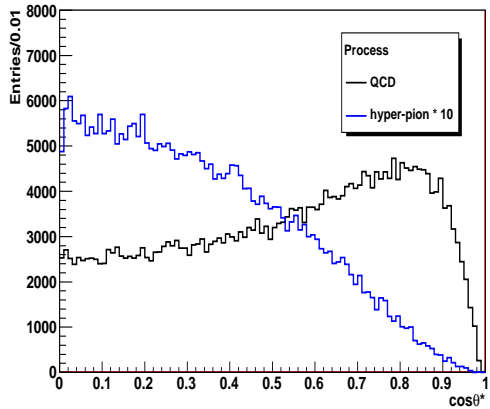}
  \caption{Distribution of the reconstructed scattering angle in the 
    center-of-mass frame of the reconstructed four highest-$p_T$ jets. 
    Note, for this shape comparison the signal was scaled by a factor of $10$.}
  \label{fig:CosThetaMarginal}
\end{figure}

To improve further the rejection of the background, the relative difference between the two 
reconstructed masses is required to agree within $7.5$\%, i.e. $|M_1 - M_2|/(M_1 + M_2)<0.075$.
Additional separation between background and signal is obtained by using the reconstructed 
scattering angle ($\cos(\theta^*)$) in the center-of-mass frame of the four jets to be 
less than~$0.6$. The $\cos(\theta^*)$ distribution is shown in Figure~\ref{fig:CosThetaMarginal} 
for the hyper-pion signal and the QCD background. All cuts but the cut on the scattering angle 
were applied. For the signal the distribution is $\sin(2\theta^*)$ like as the production of 
scalar particles is central, whereas for the  QCD background the production in the forward 
region is more pronounced. 

\begin{figure}[!t]
  \centering
  \includegraphics[width=7cm,height=6cm]{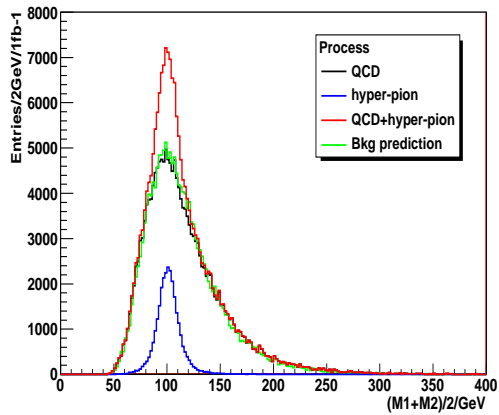}
  \caption{The average mass for a hyper-pion signal of $m_\ourpi=100 \Gev$, the QCD background 
    and the sum of signal plus background are shown after all cuts for an integrated luminosity 
    of $1{\rm fb}^{-1}$. The green line is the predicted background according to the ABCD method.}
  \label{fig:SignalandBG}
\end{figure}

\begin{figure}[!t]
  \centering
  \includegraphics[width=6cm]{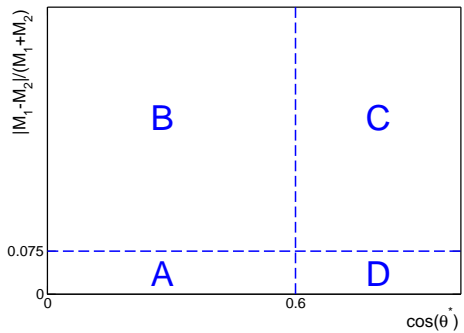}
  \caption{The four event categories used in the ABCD method. The two discriminators are the
    scattering angle $\cos(\theta^*)$ and the relative mass difference of the two reconstructed 
    resonance candidates. The signal-enriched region corresponds to selection A.}
  \label{fig:ABCD}
\end{figure}

The mean of the two reconstructed masses for events that pass all cuts is shown in 
Figure~\ref{fig:SignalandBG} for a hyper-pion signal of $m_{\ourpi}=100 \Gev$. The data 
corresponds to an integrated luminosity of $1{\rm fb}^{-1}$ at the $7 \Tev$ LHC. As the 
form of the signal is similar to the form of the background, the ABCD matrix method is 
used to predict the background. The four regions are separated via the scattering 
angle (less than or greater than $0.6$) and the relative mass difference (inverting the cut 
described above), cf. Figure~\ref{fig:ABCD}. The region with the highest signal purity
for that we want to estimate the background normalisation and shape is denoted as A. 
Under the assumption our two discriminators are uncorrelated for the QCD background, we 
can estimate the number of background events in region A to 
\beq
N^A_{\rm BG} = \frac{N^B_{\rm BG}N^D_{\rm BG}}{N^C_{\rm BG}}\,.
\eeq
The background shape we take from region $D$, where only the cut on the scattering angle 
is reversed with respect to the signal selection. The background prediction from the ABCD 
method is shown in Figure~\ref{fig:SignalandBG}. Both the background shape as well as the 
normalisation are predicted well when comparing to the Monte-Carlo truth. This data driven 
approach will have to be validated with actual collision data of course, i.e., will the 
same correlation pattern be observed in the data as predicted by Monte Carlo and which 
scale factor will have to be applied to the Monte-Carlo prediction to account for 
higher-order corrections. Here our emphasis was on the background per se, in the actual 
data potential signal contaminations in regions $B$, $C$ and $D$ will have to be taken 
into account as well. 

\section{Results}
\label{sec:lhc_potential}

\begin{figure}[!t]
  \centering
  \includegraphics[width=7cm,height=6cm]{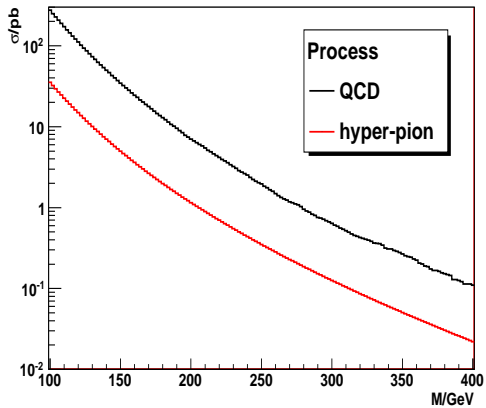}
  \caption{Cross sections after all cuts for the QCD background,
    and the hyper-pion signal.}
  \label{fig:CrossSectionAllCuts}
\end{figure}

In Figure~\ref{fig:CrossSectionAllCuts} the cross sections after all cuts are shown 
as function of the scalar mass. The QCD background decreases strongly as expected. 

\begin{figure}[!t]
  \centering
  \includegraphics[width=7cm,height=6cm]{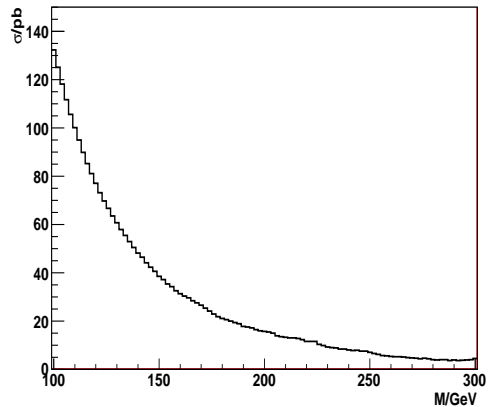}
  \caption{The sensitivity of the analysis is shown, based on the 
    statistical error in a window of $40 \Gev$ centered on the nominal 
    mass.}
  \label{fig:LHCXSectionLimit}
\end{figure}

In Figure~\ref{fig:LHCXSectionLimit} the sensitivity of the analysis is shown as function 
of the hyper-pion mass. We define the sensitivity as the statistical error on the background 
(and its prediction) for $1{\rm fb}^{-1}$ in a window of $40 \Gev$ centered on the nominal 
hyper-pion mass divided by the efficiency and multiplied by $1.64$ (similar to a one-sided 
Gaussian counting experiment limit).

As the instantaneous luminosity increases the trigger thresholds of the experiments will 
increase, using as we are using a sliding $p_T$ cut, cf. Sec.~\ref{sec:evt_selection}, the 
analysis is valid for masses greater than twice the trigger threshold. Lower masses than 
that are not accessible. For masses far from the trigger threshold one could perform a 
sideband analysis only in the signal region (bump hunting), i.e., an analysis with a fixed 
cut on the transverse momentum. This however can lead to double peak structures for the signal, 
diluting somewhat the gain on the efficiency.

For a mass of $300 \Gev$ the hyper-pion (sgluon) cross section is $10$~pb ($13$~pb). The 
sensitivity of the analysis is $5$~pb. Thus even adding experimental errors, the unknown 
K-factor of the background cross section, we can surely state that the four-jet final state 
can be probed in spite of the huge QCD background.

\begin{figure}[!t]
  \centering
  \includegraphics[width=7cm,height=6cm]{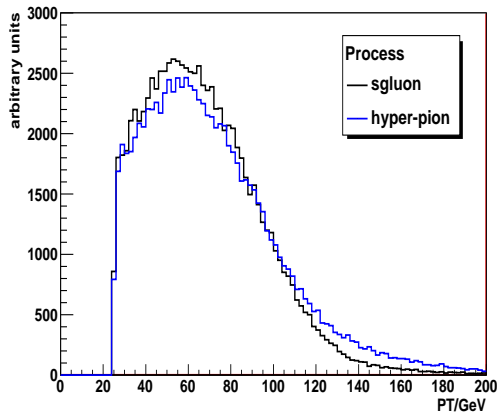}
  \caption{The distribution of the $4^{\rm th}$ jet is shown for the sgluon and 
    hyper-pion signal for a scalar mass of $300 \Gev$ (before cuts).}
  \label{fig:PT300}
\end{figure}

The difference between the hyper-pion and sgluon pair production goes beyond a scaling 
of the cross sections. The $s$-channel coloron exchange also has an impact on the 
kinematics and therefore on the selection efficiency. 
In Figure~\ref{fig:PT300} the transverse-momentum distribution of the $4$th jet is shown
for a signal mass of $300 \Gev$. The $s$-channel coloron increases the $p_T$ and therefore 
for the same analysis the efficiency for hyper-pions and sgluons is not the same as shown 
in Table~\ref{tab:eff}. While the efficiency for the hyper-pions is stable, the efficiency 
for sgluons decreases slowly for increasing masses. It is interesting to note, that the 
efficiency as a function of the $p_T$ cut behaves differently in the two scenarios. For example 
for a scalar mass of $500 \Gev$ a decrease of the $p_T$ cut by $20\%$ leads to an increase
of the hyper-pion selection efficiency by a factor of $2$, the efficiency of the sgluon
signal, however, increases by a factor of $3$. A dedicated additional analysis for 
masses far from the trigger threshold could be envisaged that could provide further information 
to discriminate the two models.

\begin{table}[htb]
\begin{center}
\begin{tabular}{lcccc}
           & $100 \Gev$ & $225 \Gev$ & $300 \Gev$ & $500 \Gev$ \\ 
\hline
hyper-pion & $0.9\%$  & $1.1\%$ & $1.2\%$ & $1.2\%$ \\ 
sgluon     & $0.6\%$  & $0.4\%$ & $0.3\%$ & $0.2\%$ \\
\end{tabular}
\end{center}
\caption{\label{tab:eff} Efficiencies are shown as function of the scalar mass.}
\end{table}

\begin{figure}[!t]
  \centering
  \includegraphics[width=4.9cm]{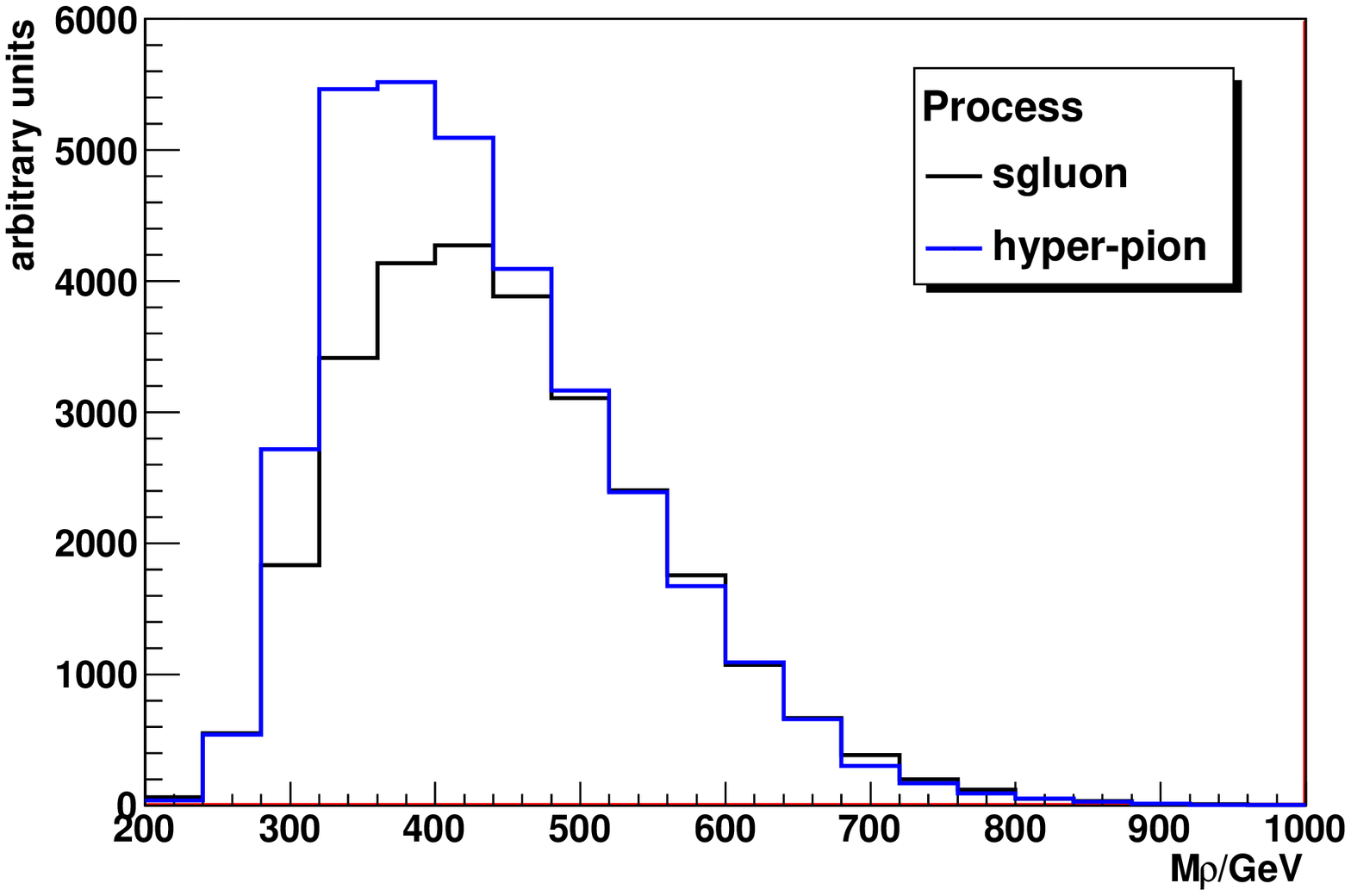}
  \includegraphics[width=4.9cm]{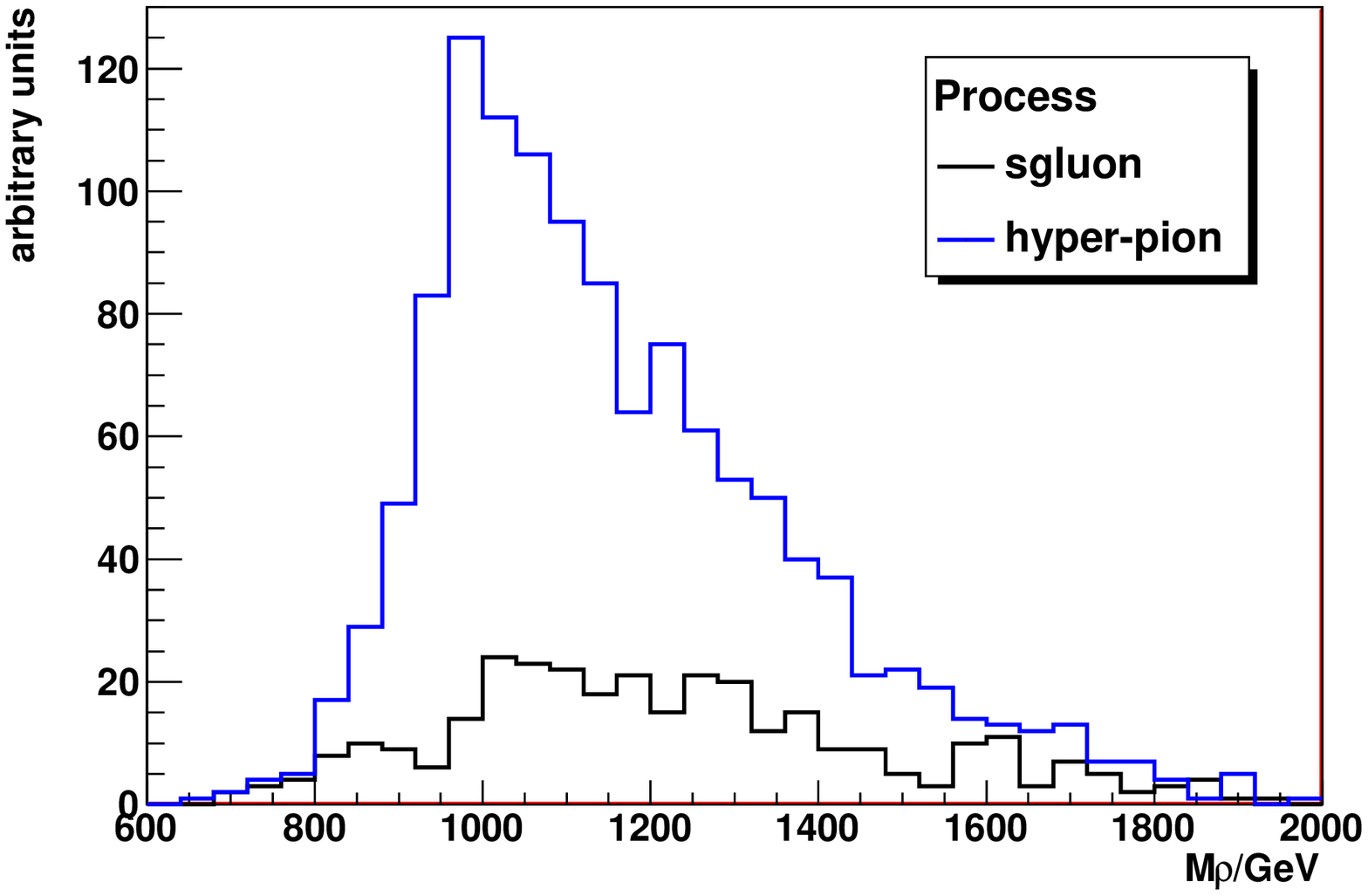}
  \includegraphics[width=4.9cm]{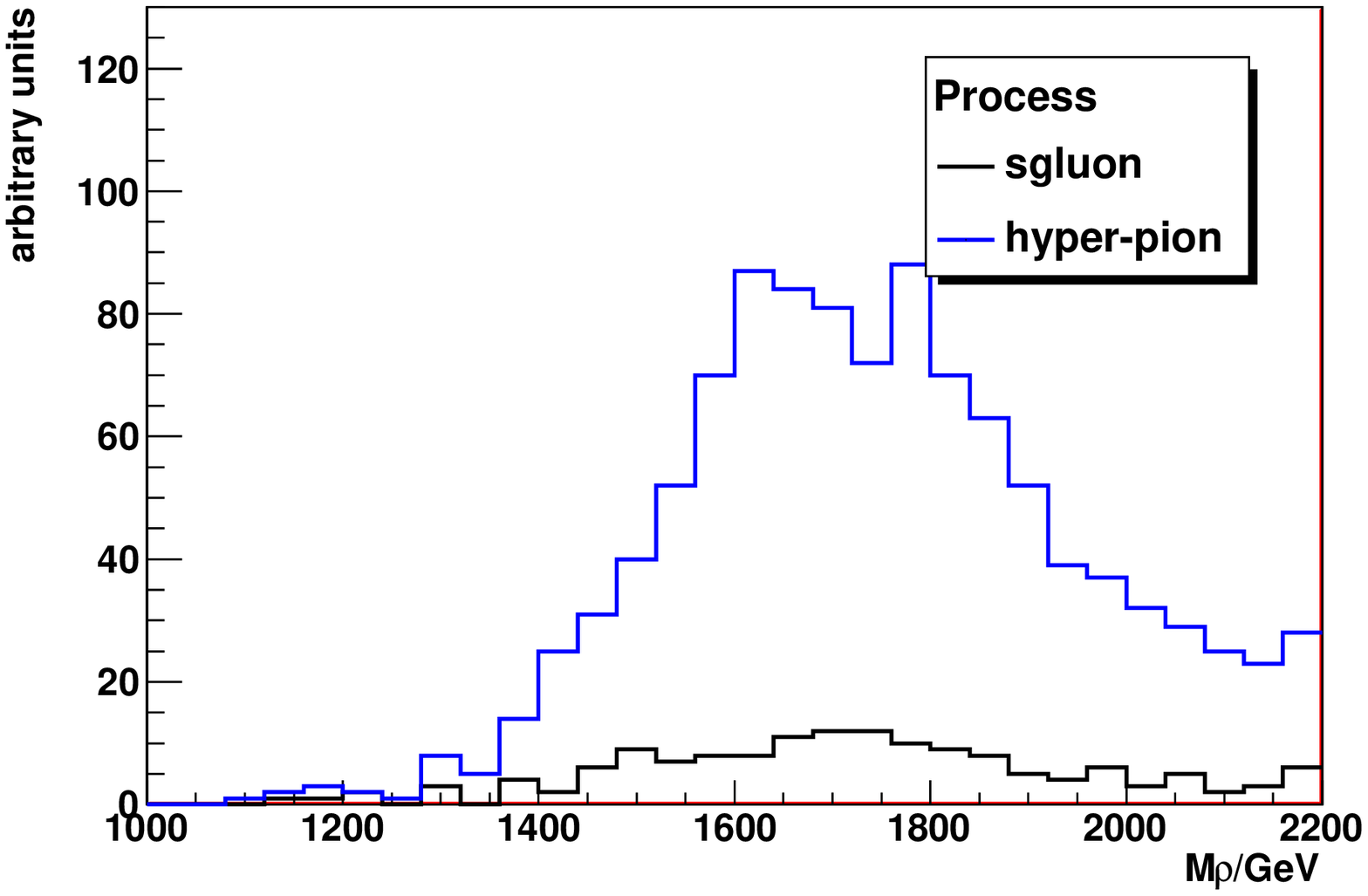}
  \caption{The distribution of the reconstructed coloron mass is shown 
    after all cuts for the hyper-pion model and the sgluon model. For 
    the latter the distribution is flatter, whereas for the former the 
    $s$-channel contribution is visible. From left to right the distributions 
    correspond to scalar masses of $100$, $300$ and $500 \Gev$. In the 
    hyper-color scenario the corresponding coloron masses are $333$, $1000$ and
    $1667 \Gev$, respectively.}
  \label{fig:Coloron}
\end{figure}

\noindent
An interesting question arises also whether the coloron mass can be reconstructed. To 
address this question we consider the invariant mass of the two reconstructed scalar 
candidates. As can be seen in Figure~\ref{fig:Coloron} at low mass the coloron mass 
distribution is diluted by the non-resonant production mechanism. No clear distinction 
can be observed between the sgluon and hyper-pion case. However for masses of $300 \Gev$ 
as well as $500 \Gev$, cf. the middle and right panel of Figure~\ref{fig:Coloron}, the 
coloron of the hyper-pion model becomes visible, thus giving us possibly a handle to 
distinguish the two models. We note however that in order to differentiate the two 
hypotheses, the actual QCD background (larger than the signal) will have to be dealt with. 
While the signal differences are quite striking, an integrated luminosity much larger 
than $1 {\rm fb}^{-1}$ will be necessary to conclude on the underlying model and only if 
the signal turns out to be at sufficiently high mass. However, the simulation of such large 
integrated luminosities for the background is beyond the scope of our paper. Furthermore 
we want to add that for an LHC center-of-mass energy of $14 \Tev$ the discrimination of 
the two models through the reconstruction of the coloron resonance gets more complicated due
to the different initial-state parton fluxes. For the hyper-pion mass range considered here, 
namely $m_\ourpi = 100 ... 500 \Gev$, at $14 \Tev$ the hyper-pion to sgluon cross-section 
ratio is strictly smaller than for $7 \Tev$. In turn the relative importance of the coloron 
contribution to the total hyper-pion pair-production cross section is largely reduced and 
the QCD continuum production mechanism dominates. On the other hand this dependence on the
collision energy might serve as a handle to establish the coloron production
from $q\bar q$ initial states.

The alternative way to verify the existence of the hyper-rho state is to study its 
pair-production process. In refs.~\cite{Kilic:2008ub,Dicus:2010bm,Sayre:2011ed} the 
signature $pp \to \col\col \to 4\ourpi\to 8 {\rm jets}$ has been studied and it was 
shown that the LHC provides great discovery potential both at $7$ and $14 \Tev$ 
center-of-mass energy. 

\section{Conclusions}
\label{sec:conclusions}

We have presented a search strategy for the pair production of new color-adjoint scalar 
particles that decay into pairs of jets. Such states are predicted by many extensions of 
the Standard Model. New color octets can be produced in abundance at the LHC and their 
production rate is rather model independent. We considered two benchmark models as 
predictive realisations, a simplified sgluon model and a vectorlike confinement scenario 
that features an additional vector resonance in the color adjoint. 

While the QCD backgrounds seem overwhelming at first sight, the use of an equal-mass 
constraint for the two reconstructed dijet resonances and the angular distribution of 
the production of the scalars help to reduce the backgrounds to a manageable level. 
Both models are clearly detectable at the LHC for a broad range of masses. When 
searching for scalar octets with $m_{S_8}\gsim 300 \Gev$ such that $m_{S_8}/2$ is well 
above the trigger threshold the analysis cuts might be adjusted to further optimise the 
signal yield and increase statistics. Comparing the two benchmark models for different 
masses at low mass only the cross section provides some differentiation. If the scalars 
have higher mass, the reconstruction of the four-jet invariant mass might shed light on 
the presence of an additional $s$-channel resonance decaying into a pair of scalars, 
however an integrated luminosity bigger than $1 {\rm fb}^{-1}$ is needed.


\begin{acknowledgments}
We thank Peter Zerwas for helpful discussions on the sgluon model. SS 
would like to thank Can Kili\c{c} for fruitful discussions and 
Elmar Bittner for computing support. Most simulations underlying this 
study have been performed on bwGRiD (\url{http://www.bw-grid.de}), member 
of the German D-Grid initiative, funded by the Ministry for Education and
Research and the Ministry for Science, Research and Arts Baden-W\"urttemberg.
Part of these studies were performed in the context of the GDR Terascale (CNRS).
\end{acknowledgments}


\appendix
\section{Cross section ratio}

The ratio of the cross sections for the pair production of hyper-pions and sgluons is given
in Table~\ref{tab:ratioHyperPionSgluon}. The calculation was performed at leading
order using the CTEQ6L1 PDFs \cite{Pumplin:2002vw}.

\begin{table}[htb]
\begin{center}
\begin{tabular}{cc|cc|cc|cc}
M/GeV & & M/GeV & & M/GeV & & M/GeV & \\
100 & 0.6 & 330 & 0.79 & 560 & 0.84 & 790 & 0.74 \\
110 & 0.6 & 340 & 0.8 & 570 & 0.84 & 800 & 0.73 \\
120 & 0.61 & 350 & 0.8 & 580 & 0.84 & 810 & 0.73 \\
130 & 0.62 & 360 & 0.81 & 590 & 0.84 & 820 & 0.72 \\
140 & 0.63 & 370 & 0.81 & 600 & 0.83 & 830 & 0.71 \\
150 & 0.64 & 380 & 0.82 & 610 & 0.83 & 840 & 0.71 \\
160 & 0.65 & 390 & 0.82 & 620 & 0.83 & 850 & 0.7 \\
170 & 0.66 & 400 & 0.83 & 630 & 0.82 & 860 & 0.69 \\
180 & 0.67 & 410 & 0.83 & 640 & 0.82 & 870 & 0.69 \\
190 & 0.68 & 420 & 0.83 & 650 & 0.82 & 880 & 0.68 \\
200 & 0.69 & 430 & 0.84 & 660 & 0.81 & 890 & 0.67 \\
210 & 0.7 & 440 & 0.84 & 670 & 0.81 & 900 & 0.66 \\
220 & 0.71 & 450 & 0.84 & 680 & 0.8 & 910 & 0.66 \\
230 & 0.72 & 460 & 0.84 & 690 & 0.8 & 920 & 0.65 \\
240 & 0.72 & 470 & 0.84 & 700 & 0.79 & 930 & 0.64 \\
250 & 0.73 & 480 & 0.84 & 710 & 0.79 & 940 & 0.64 \\
260 & 0.74 & 490 & 0.85 & 720 & 0.78 & 950 & 0.63 \\
270 & 0.75 & 500 & 0.85 & 730 & 0.78 & 960 & 0.62 \\
280 & 0.76 & 510 & 0.85 & 740 & 0.77 & 970 & 0.62 \\
290 & 0.76 & 520 & 0.85 & 750 & 0.76 & 980 & 0.61 \\
300 & 0.77 & 530 & 0.85 & 760 & 0.76 & 990 & 0.6 \\
310 & 0.78 & 540 & 0.85 & 770 & 0.75 & 1000 & 0.6 \\
320 & 0.78 & 550 & 0.84 & 780 & 0.75 &  &   \\
\end{tabular}
\end{center}
\caption{\label{tab:ratioHyperPionSgluon} Ratio of the hyper-pion 
  cross section to the sgluon cross section.}
\end{table}


\end{document}